\documentclass[conference]{IEEEtran}
\IEEEoverridecommandlockouts
\usepackage{cite}
\usepackage{amsmath,amssymb,amsfonts}
\usepackage{algorithmic}
\usepackage{graphicx}
\usepackage{textcomp}
\usepackage{xcolor}
\usepackage{url}
\def\BibTeX{{\rm B\kern-.05em{\sc i\kern-.025em b}\kern-.08em
    T\kern-.1667em\lower.7ex\hbox{E}\kern-.125emX}}
\begin{document}

\title{Ransomware Detection and
	Classification using Machine
	Learning\\
}
\author{\IEEEauthorblockN{Kavitha Kunku}
\IEEEauthorblockA{\textit{Dept. of Physics \& Computer Science} \\
\textit{Wilfrid Laurier University (WLU)}\\
Waterloo, ON, Canada \\
kunk0720@mylaurier.ca}
\and
\IEEEauthorblockN{ANK Zaman}
\IEEEauthorblockA{\textit{Dept. of Physics \& Computer Science} \\
	\textit{Wilfrid Laurier University (WLU)}\\
	Waterloo, ON, Canada \\
azaman@wlu.ca}
\and
\IEEEauthorblockN{Kaushik Roy}
\IEEEauthorblockA{\textit{Dept. of Computer Science} \\
\textit{North Carolina A\&T State University}\\
Greensboro, NC, USA \\
 kroy@ncat.edu}\\
}

\maketitle

\begin{abstract}
Vicious assaults, malware, and various ransomware pose a cybersecurity threat, causing considerable damage to computer structures, servers, and mobile and web apps across various industries and businesses. These safety concerns are important and must be addressed immediately. Ransomware detection and classification are critical for guaranteeing rapid reaction and prevention. This study uses the XGBoost classifier and Random Forest (RF) algorithms to detect and classify ransomware attacks. This approach involves analyzing the behaviour of ransomware and extracting relevant features that can help distinguish between different ransomware families. \\
The models are evaluated on a dataset of ransomware attacks and demonstrate their effectiveness in accurately detecting and classifying ransomware. The results show that the XGBoost classifier, Random Forest Classifiers, can effectively detect and classify different ransomware attacks with high accuracy, thereby providing a valuable tool for enhancing cybersecurity.

\end{abstract}

\begin{IEEEkeywords}
	Ransomware, Malware, Malware Detection, Classification, XGBoost, Random Forest, Cybersecurity
\end{IEEEkeywords}

\section{Introduction}\label{sec1}
Ransomware attacks are becoming increasingly common and devastating for both individuals and organizations. The primary objective of this research work is to develop a Machine-Learning (ML) based system that can detect ransomware attacks in real-time and classify them into different categories~\cite{salwa23},\cite{int1}. The system will be designed to analyze the behaviour of malicious software and identify the specific type of ransomware being used. This information will improve the system's accuracy and provide relevant information to victims and cybersecurity professionals\cite{int2}. \\
Ransomware is malicious software that encrypts the user's files or entire system, making it inoperable, and then demands a ransom fee from the victim's computer in exchange for the decryption key \cite{int2}. Ransomware attacks have become more common in recent years, causing severe financial and reputational harm to individuals and organizations \cite{int2}.\\
Detecting and classifying ransomware is an important activity for cybersecurity professionals to perform to safeguard against assaults \cite{int3}. Detecting and classifying ransomware entails recognizing its behaviour and features, separating it from different malware, and discovering its origin and attack pathways.\\
Machine learning in identifying ransomware is an emerging research subject with tremendous application potential in creating anti-ransomware systems \cite{int3}. Malware, including ransomware, can be detected immediately through its unpredictable actions using methods based on machine learning, enhancing security. Using machine learning, computers can learn and find patterns in enormous volumes of data and generate predictions based on those patterns \cite{int4}. In the context of ransomware detection and classification, machine learning can be used to analyze various features of ransomware attacks, such as the type of encryption used, the attack vector, and the behaviour of the ransomware.\\
This research explores machine learning techniques to detect and classify ransomware attacks \cite{int4, mowri2022application}. Specifically, the proposed technique can analyze various features of ransomware attacks and predict whether a given attack is ransomware. This research will also explore different machine learning algorithms like XGBoost and Random Forest for classifying the ransomware and evaluate their performances on the ransomware detection and classification task.

The rest of the paper is organized as follows: section~\ref{sec2} surveys and presents recently published
work related to this work, section~\ref{sec3} describes the dataset used in the research, section~\ref{sec4} represents the technical details of the implementation, section~\ref{sec5} represents results, and section~\ref{sec6} describes the concluding remarks of this paper.


\section{Related Work}\label{sec2}
A significant amount of research has been done in the area of ransomware detection and classification using machine learning. This literature survey briefly reviews some of the recent and notable works in this field. Conventional detection methods have been used to classify various threats, including ransomware. \\A clearly defined behavioural framework may be utilized to examine different malware families, and many malware families have common behavioral traits such as payload persistence, stealth methods, and network activity\cite{lr2}. The most often used conventional malware protection system is signature-based analysis, and A. M. Abiola and M. F. Marhusin\cite{lr2} proposed a signature-based detection model for malware by extracting the Brontok worms, and an n-gram technique was utilised to break down the signatures that were extracted.\\
The framework detects malware and generates an accurate solution which removes all dangers. To address the problem, a combination of a static and dynamic based or behaviour-based structure was developed\cite{lr3}, in which analysis static-based method analyses the application's code to identify illicit operations and dynamic based analysis monitors the processes to determine the behaviour of malicious users and will be identified as suspicious and subsequently terminated.\\
Another notable work in this field was by Singh et al. (2019) \cite{lr1}, who proposed a system called Ransom Detector, which used a random forest classifier to detect and classify ransomware. Ransom Detector analyzed various features of the ransomware, including file system activity, network traffic, and process information\cite{lr1}. The system achieved an accuracy of 99.6\% in detecting ransomware attacks. Random Forest is a popular machine learning technique that is utilized for identifying malware and ransomware. F. Khan et al.,~\cite{lr4} developed a DNAact-Ran-based malware detection technique based on sequencing design constraints and the k-mer frequency vector. The structure was tested on 582 DNAact-Run ransomware samples and 942 legitimate  instances to assess precision, recall, f-measure, and accuracy.
S. Poudyalwe et al.\cite{lr5} developed a machine learning-based detection algorithm for identifying malware which employs multiple levels analysis for better understanding the intent of malware code parts. The algorithm was tested, and the findings show that it detects malware with a 76\% to 97\% accuracy.\\
To summarize, machine learning has been widely utilised to detect and classify ransomware assaults\cite{lr5}. To train machine learning algorithms, many features such as network traffic, file system activity, and process information have been analysed. The findings of these studies indicate that machine learning can be an effective technique in tackling the ransomware problem.

\section{Dataset}\label{sec3}
\begin{figure}[ht]
	\includegraphics[scale=.7]{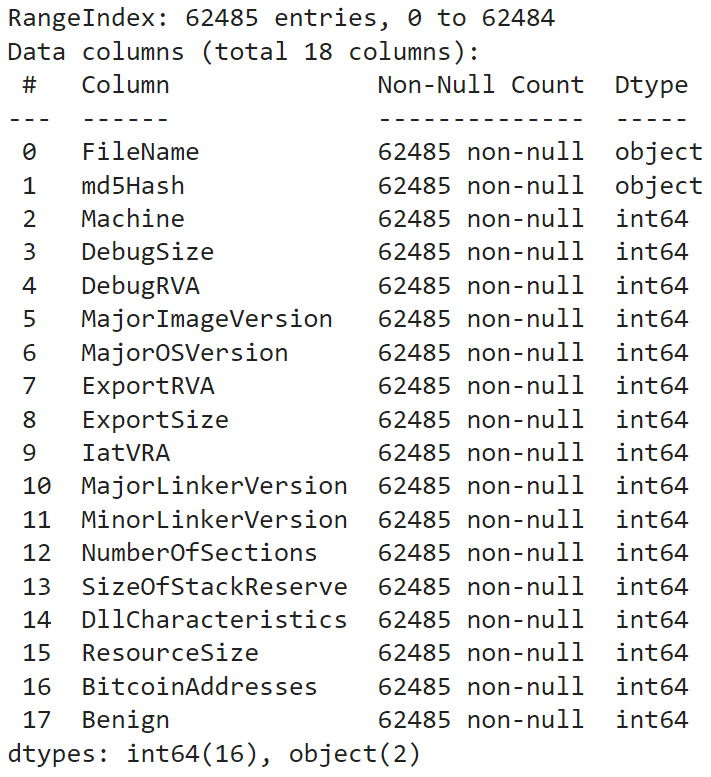}
	\centering
	\caption{Meta data: Ransomware Detection Dataset }
	\label{fig:metadata}
\end{figure}

This research uses the ransomware dataset~\cite{dataset1, dataset2} from Kaggle. The dataset consists of 62485 unique values and 18 features. The meta data of Ransomware Detection dataset is presented in Figure~\ref{fig:metadata}.

\begin{figure}[ht]
	\includegraphics[scale=1]{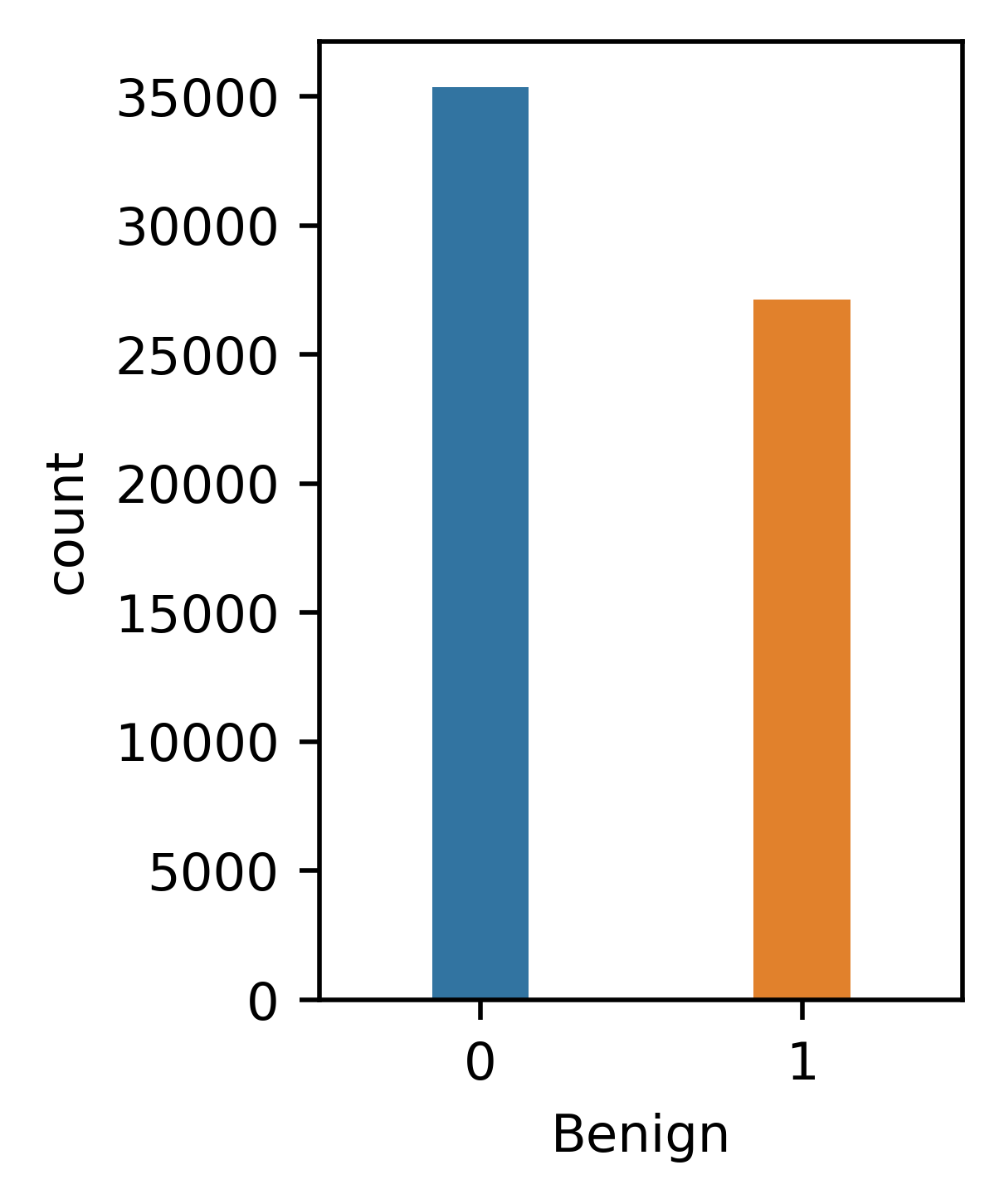}
	\centering
	\caption{Legitimate (1) VS Malware (0) Data }
	\label{fig:01data}
\end{figure}

The dataset has 27118 legitimate collections and the remaining 35367 are malicious collections. Figure~\ref{fig:01data} shows the ratio of the legitimate and non-legitimate data portions.

The selected dataset concentrates more on the changes that the system undergoes when it has been attacked with any malware in terms of Debug Size, which is the size of the debug directory; Major Image version, which is the version of the file; Major OS version, Export RVA, Export Version, IatRVA is the virtual address of the import file.

\begin{figure}[ht]
	\includegraphics[scale=.28]{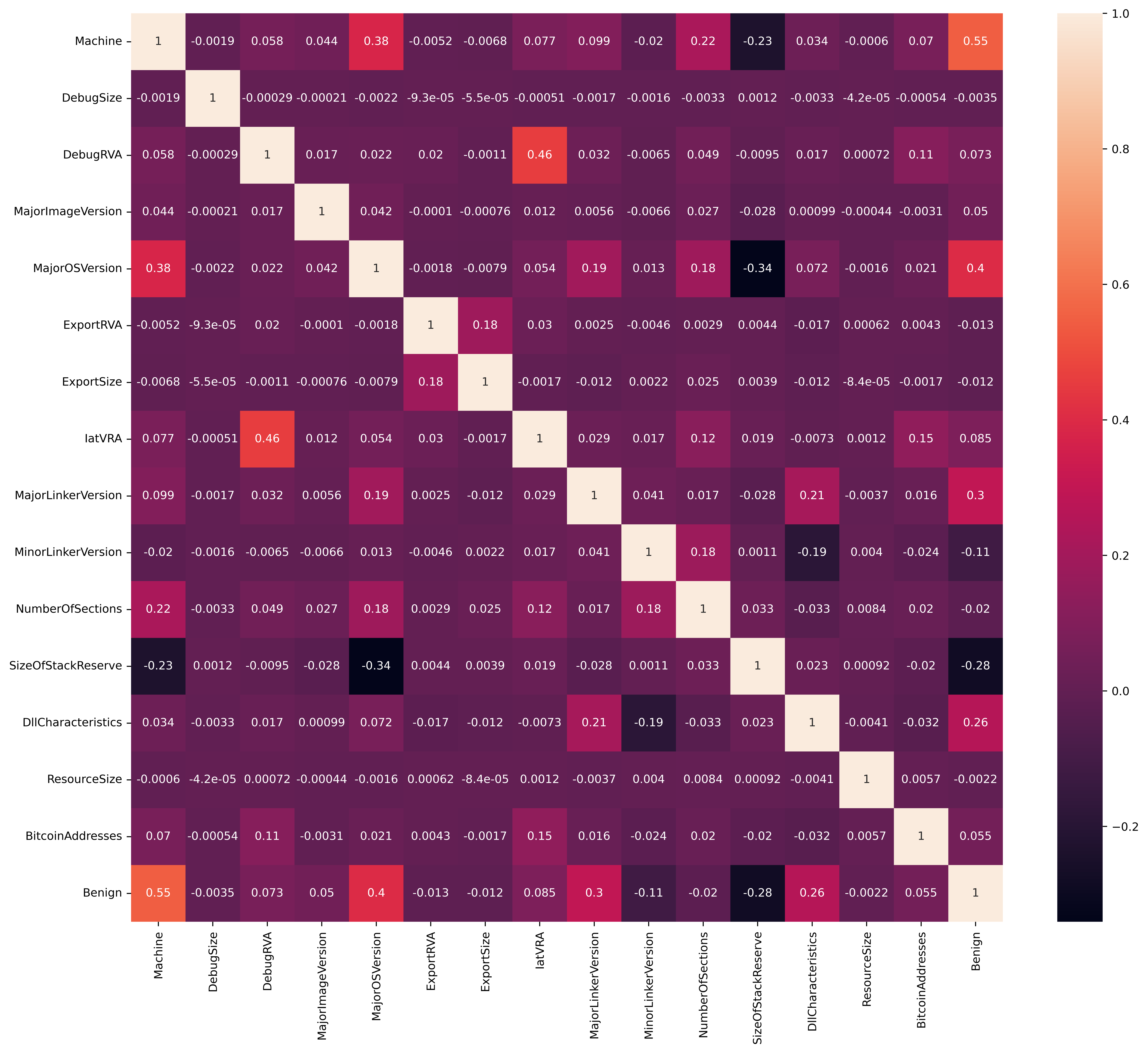}
	\centering
	\caption{Correlation Matrices Among Dataset Features }
	\label{fig:htdata}
\end{figure}

Using a heat map, Figure~\ref{fig:htdata} represents the correlation matrices among data features. The diagonal boxes are all bright and have values that correlate with each feature. The linear trend between two features is defined by color hues ranging from light to dark. 

Table~\ref{tab:data}, below shows the changes observed if a system or file malfunctions or is corrupted.

\begin{table}[!ht]
	\centering
	\scriptsize	
	\caption{Data set Observations}\label{dataset}
	\label{tab:data}
	\resizebox{.49\textwidth}{!}{
	\begin{tabular}{|c|c|c|c|c|}
		\hline
		Data & MajorImage Version & Export Size & IatRVA & MinResourceSize  \\ \hline
		Legitimate  & Non Zero & Non Zero& 4096& Non Zero \\ \hline
		Malware &0 &0 &0 or very large number &0 \\ \hline
	\end{tabular}
}
\end{table}

\section{Implementation}\label{sec4}
After surveying recently published research papers, this research has decided to carry out experiments on both Xgboost~\cite{yang2020, xgb} and Random Forest~\cite{Liu2012NewML, rf} methods for ransomware detection and classification. Xgboost is a popular gradient-boosting algorithm that has been used in various machine learning applications, including malware detection \cite{lr5}. It is an ensemble machine learning algorithm that combines weak and strong learners to produce a learner who is powerful. The XGBoost classifier is a subset of the XGBoost algorithm that is optimized for classification issues.
\subsection{\textbf{Algorithms}}
\begin{enumerate}
	\item \textbf{XGBoost Classifier:} The XGBoost ~\cite{yang2020, xgb} classifier works by constructing a series of decision trees, each of which attempts to rectify the flaws of the previous tree. The approach provides greater weights to incorrectly classified data throughout the training process, so subsequent trees focus on accurately identifying those samples\cite{other}. The final forecast takes the weighted average of all the trees' projections. The XGBoost classifier has several advantages over other classification algorithms, including high accuracy, scalability, flexibility, and speed.
	\item \textbf{Random Forest Algorithm:} Random Forest~\cite{Liu2012NewML, rf} is a renowned and widely used method among data scientists. Random forest is a supervised machine learning algorithm commonly used in problems with classification and regression \cite{other}. It constructs decision trees from several samples and uses their majority vote for classification and the average for regression. One of the essential characteristics of the Random Forest Method is that it can handle data sets with continuous variables in a regression and categorical variables for classification.
\end{enumerate}

\subsection{\textbf{Model Training and Evaluation}}
The implementation of ransomware detection and classification using Xgboost and Random Forest involves the following steps:
\begin{figure}[h]
	\includegraphics[scale=.5]{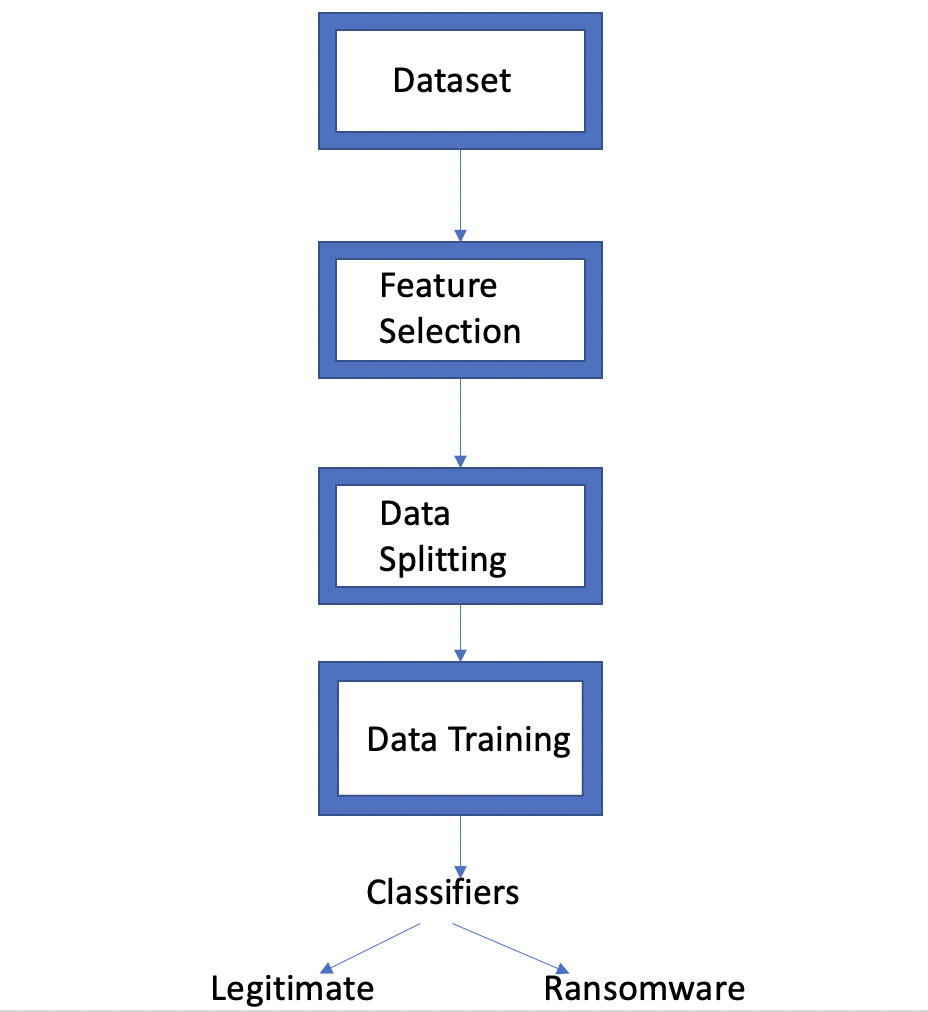}
	\centering
	\caption{Framework for Ransomware Classification Implementation}
\end{figure}

\begin{enumerate}
	\item \textbf{Data Collection:} As mentioned in the section~\ref{sec3}, the data has been collected from Kaggle~\cite{dataset1} for ransomware detection and classification, with 62485 samples and 18 features.
	\item \textbf{Feature Selection:} Appropriate features has extracted from the dataset, such as file system activity, Debug Size, Export Size, Major Image Version, IatRVA, and MinResource size, to analyze the changes in the data that is affected by malware attacks and the characteristics of a normal system which has legitimate data.
	\item \textbf{Data Preprocessing:} The selected dataset for this research is clean and clear. It has no null or missing values in the data set. So it did not require any preprocessing steps for this data.
	\item \textbf{Data Splitting:} The collected data has been split into training and testing datasets typically into an 80:20 ratio in which 80\%  of data is used for training the model, and 20\% is used as test data.
	\item \textbf{Model Training: } XGBoost, and Random Forest classifiers are trained using the Scikit-learn~\cite{sl} python library. During the training phase, these algorithms construct a chain of decision trees, with each succeeding tree attempting to fix the flaws of the prior tree.
	\item \textbf{Model Evaluation:} The performance of both algorithms is evaluated by calculating evaluating metrics like accuracy scores, Precision, Recall, and F1 scores as described below by testing the data against a test data set which was not used during the training process.\\
	\textbf{Precision:} Mathematically, Precision is defined as the ratio of true positives to the sum of true positives and false positives (FP).
		
	\begin{equation}
		\text{Precision} = \frac{\text{True Positives}}{\text{True Positives} + \text{False Positives}}
	\end{equation}
	
	\textbf{Recall: }Mathematically, recall is defined as the ratio of true positives (TP) to the sum of true positives and false negatives (FN).
	
	\begin{equation}
		\text{Recall} = \frac{\text{True Positives}}{\text{True Positives} + \text{False Negatives}}
	\end{equation}
	
	\textbf{F1 Score: }F1 score is the harmonic mean of precision and recall, and provides a single metric that balances both measures. It is defined as:
	\begin{equation}
		F_1 = 2*\frac{\text{Precision}*\text{Recall}}{\text{Precision} + \text{Recall}}
	\end{equation}
	
\end{enumerate}

\section{Result and Discussion}\label{sec5}
\subsection{\textbf{Results}}
XGBoost, and Random Forest algorithms are applied to classify legitimate and ransomware data. The below table shows the performance evaluation  metrics such as Accuracy, Precision, Recall, F1 score for both classifiers. XGboost algorithm achieved 99.61\% accuracy where as Random Forest secured overall 99.70\% accuracy. XGBoost Classifier got Precision, Recall, F1 scores are 0.9974, 0.9938, 0.9956 respectively. Random Forest Algorithm achieved 0.9974, 0.9949, 0.9962 Precision, Recall, F1 score respectively.

\begin{table}[ht]
	\centering
	\scriptsize	
	\caption{Experimental Results}
	\label{tab:result}
	\begin{tabular}{|c|c|c|c|c|}
		\hline
		Model & Accuracy & Precision & Recall & F1 Score  \\ \hline
		XGBoost &0.9961 &0.9974 &0.9938 &0.9956 \\ \hline
		Random Forest &0.9970 &0.9976 &0.9956 &0.9966 \\ \hline
	\end{tabular}
\end{table}

\begin{figure}[ht]
	\includegraphics[scale=.6]{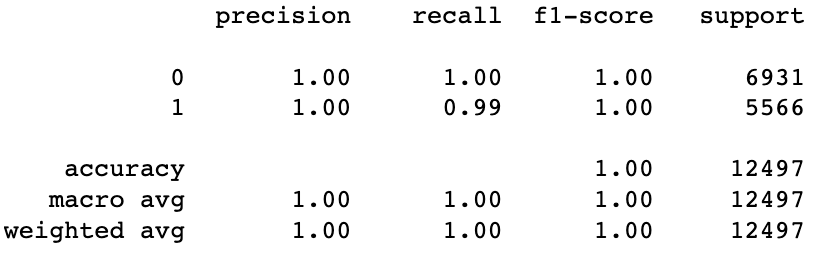}
	\centering
	\caption{Classification Report of XGBoost Classifier }
    \label{fig:xgresult}
\end{figure}

\begin{figure}[ht]
	\includegraphics[scale=.56]{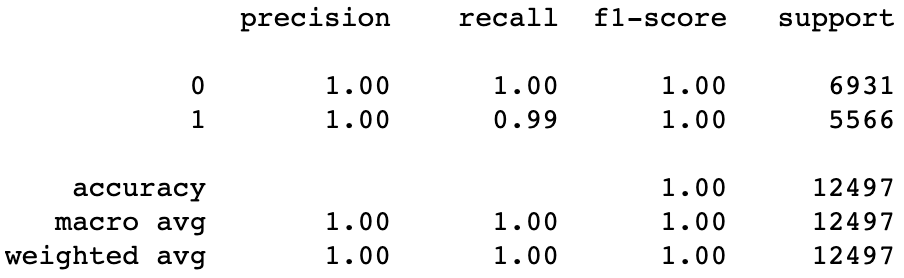}
	\centering
	\caption{Classification Report of Random Forest Algorithm }
	\label{fig:rfresult}
\end{figure}

Figure~\ref{fig:xgresult} and Figure~\ref{fig:rfresult} are showing the obtained results for both algorithms.

\begin{figure}[ht]
	\includegraphics[scale=.7]{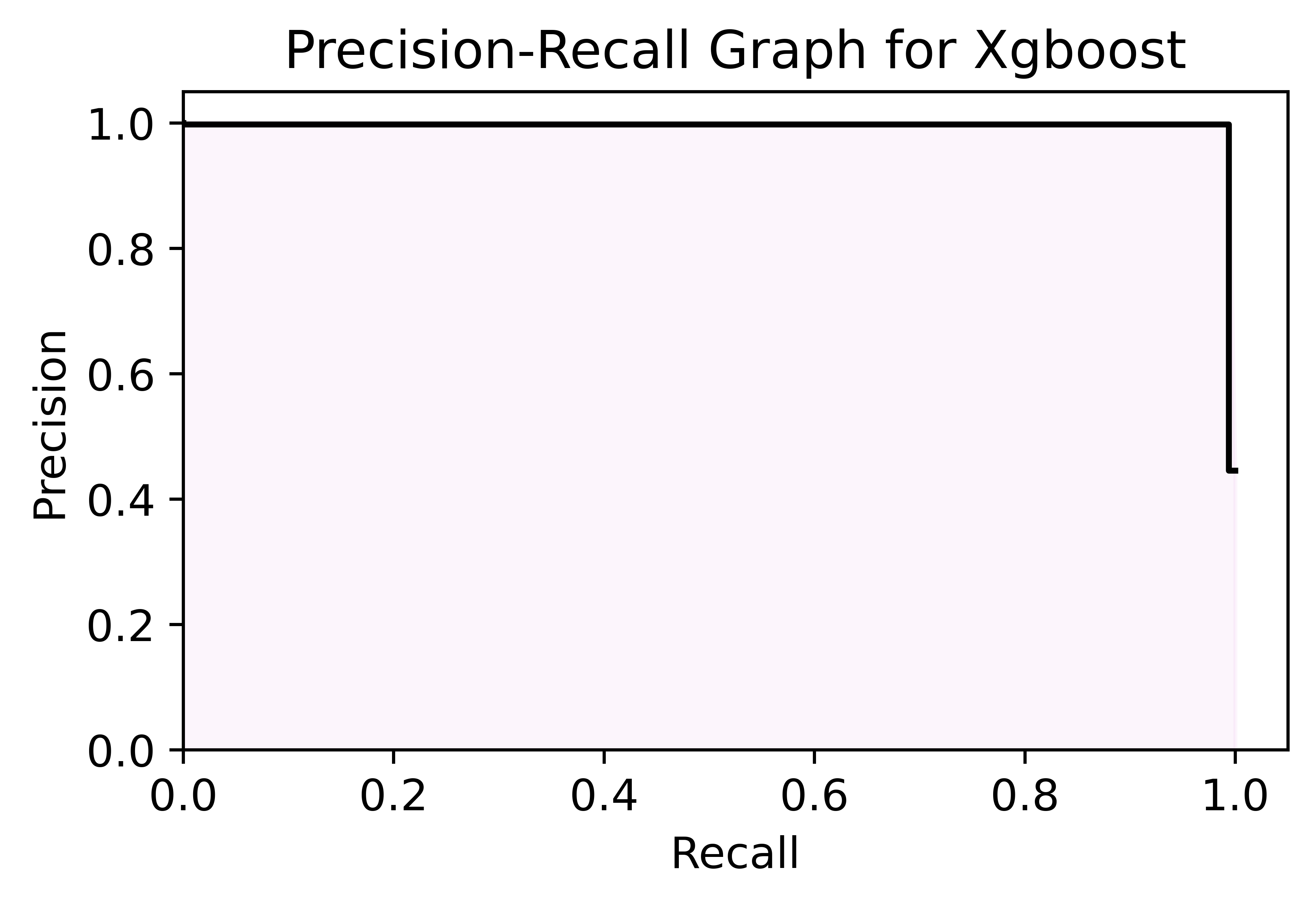}
	\centering
	\caption{Precision Recall Curve of XGBoost Classifier }
	\label{fig:prxg}
\end{figure}

\begin{figure}[ht]
	\centering
	\includegraphics[scale=.63]{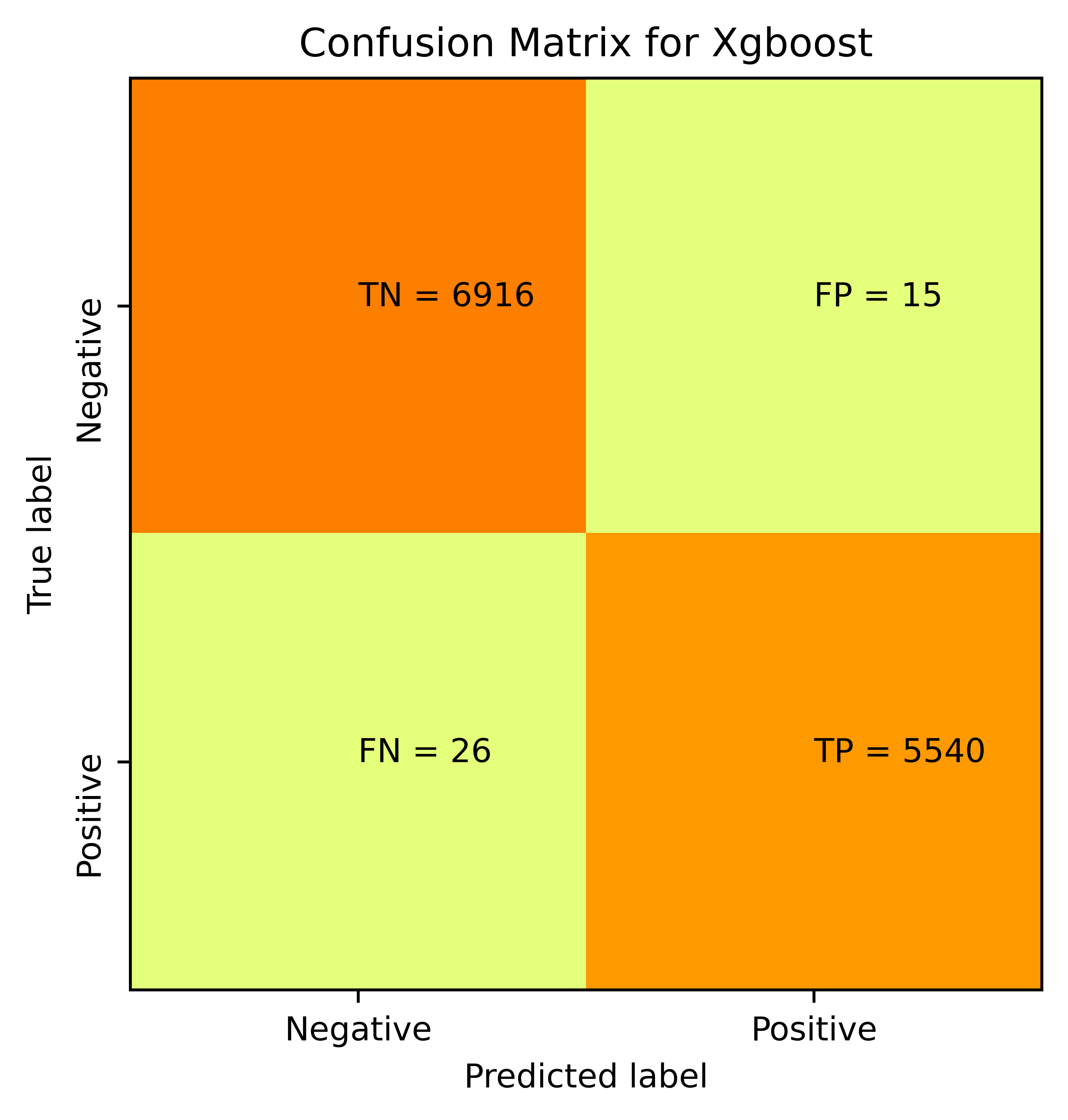}
	\caption{Confusion Matrix for XGBoost Algorithm}
	\label{fig:cmxg}
\end{figure}
\begin{figure}[!ht]
	\includegraphics[scale=.63]{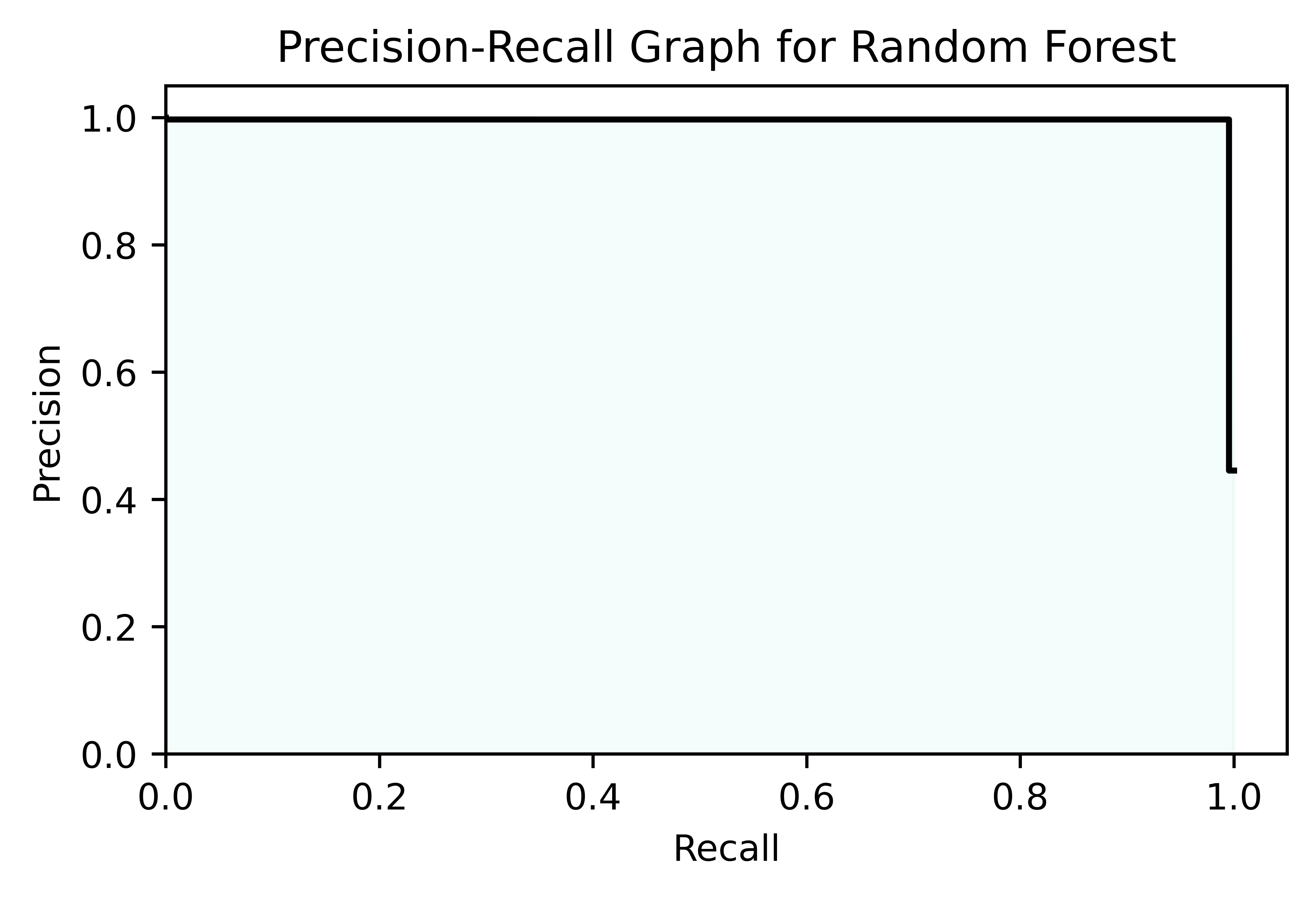}
	\centering
	\caption{Precision Recall Curve of Random Forest Algorithm }
	\label{fig:prrf}
\end{figure}

\begin{figure}[!ht]
	\centering
	\includegraphics[scale=.64]{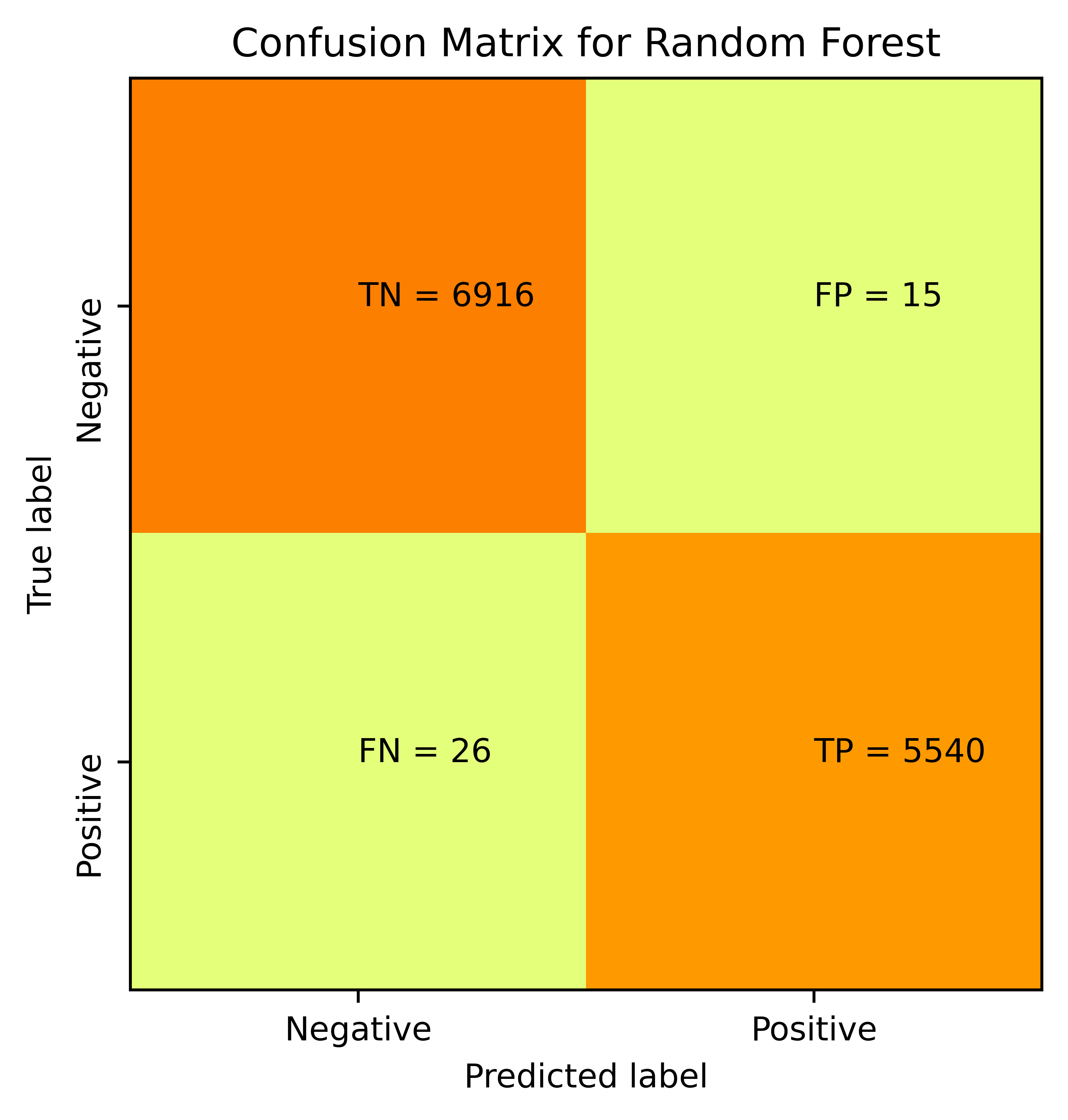}
		\caption{Confusion Matrix for Random Forest Algorithm}
		\label{fig:cmrf}
\end{figure}
Figures~\ref{fig:prxg},~\ref{fig:cmxg}~\ref{fig:prrf},~\ref{fig:cmrf}, are showing the precision-recall graphs and confusion matrices for XBboost and Randomforesr algorithms respectively.


\subsection{\textbf{Discussion}} 
The results show that both algorithms were able to achieve high accuracy in detecting and classifying ransomware samples, with Random Forest achieving a slightly higher accuracy than XGBoost which is not a noticeable change that varies in fraction of 0.09\%. 

Table~\ref{tab:rescom} shows the classification accuracies of recently published works compared to the proposed work.

\begin{table}[!ht]
	\centering
	\scriptsize	
	\caption{Comparison of results}\label{rescom}
	\label{tab:rescom}
	\resizebox{.5\textwidth}{!}{
		\begin{tabular}{|c|c|c|}
			\hline
			Reference & Classifier & Accuracy  \\ \hline
		
			Molina et al.~\cite{rica2022} & RF &94.92 \\ \hline
			Bae et al.~\cite{Bae2020RansomwareDU} &RF &98.65 \\ \hline
			Hwang et al.~\cite{10.1007/s11277-020-07166-9} &Markov Chain, RF &97.30 \\ \hline
			Rawshan et al.~\cite{mowri2022application} &Logistic
			Regression &99.15 \\ \hline
		Proposed work &XGboost, RF &99.61, 99.7 \\ \hline
			
	\end{tabular}
	}
\end{table}

This research achieved better accuracy than other works reported in Table~\ref{tab:rescom}. However, it should be noted that the performance of the models may vary depending on the data set used and the specific parameters chosen for the algorithm.


In terms of computational efficiency, the Xgboost algorithm was faster than Random Forest, which may be an advantage for large-scale data sets. However, Random Forest is known to be less prone to over fitting, which may be a concern with Xgboost.

\section{Conclusion}\label{sec6}
This research has used XGBoost, and Random Forest algorithms to classify legitimate and ransomware data. Table~\ref{tab:result} shows both classifiers' performance evaluation metrics: accuracy, precision-recall, and F1 score. XGboost algorithm achieved 99.61\% accuracy whereas Random Forest secured overall 99.70\% accuracy. XGBoost Classifier got precision, recall, and F1 scores of 0.9974, 0.9938, and 0.9956, respectively. Random Forest algorithm achieved 0.9976, 0.9956, 0.9966 precision, recall, and F1 score, respectively. In the future, the Convolutional Neural Network (CNN) based algorithms will be tested with the latest and larger scale datasets for detecting and classifying ransomware families.

\bibliographystyle{IEEEtran}
\bibliography{IEEE-CICS-Camera-2023.bib}

\begin{thebibliography}{10}
\providecommand{\url}[1]{#1}
\csname url@samestyle\endcsname
\providecommand{\newblock}{\relax}
\providecommand{\bibinfo}[2]{#2}
\providecommand{\BIBentrySTDinterwordspacing}{\spaceskip=0pt\relax}
\providecommand{\BIBentryALTinterwordstretchfactor}{4}
\providecommand{\BIBentryALTinterwordspacing}{\spaceskip=\fontdimen2\font plus
\BIBentryALTinterwordstretchfactor\fontdimen3\font minus
  \fontdimen4\font\relax}
\providecommand{\BIBforeignlanguage}[2]{{%
\expandafter\ifx\csname l@#1\endcsname\relax
\typeout{** WARNING: IEEEtran.bst: No hyphenation pattern has been}%
\typeout{** loaded for the language `#1'. Using the pattern for}%
\typeout{** the default language instead.}%
\else
\language=\csname l@#1\endcsname
\fi
#2}}
\providecommand{\BIBdecl}{\relax}
\BIBdecl

\bibitem{salwa23}
S.~Razaulla, C.~Fachkha, C.~Markarian, A.~Gawanmeh, W.~Mansoor, B.~C.~M. Fung,
  and C.~Assi, ``The age of ransomware: A survey on the evolution, taxonomy,
  and research directions,'' \emph{IEEE Access}, vol.~11, pp. 40\,698--40\,723,
  2023.

\bibitem{int1}
\BIBentryALTinterwordspacing
M.~Masum, M.~J.~H. Faruk, H.~Shahriar, K.~Qian, D.~Lo, and M.~I. Adnan,
  ``Ransomware classification and detection with machine learning algorithms,''
  in \emph{2022 {IEEE} 12th Annual Computing and Communication Workshop and
  Conference ({CCWC})}.\hskip 1em plus 0.5em minus 0.4em\relax IEEE, jan 2022.
  [Online]. Available: \url{https://doi.org/10.1109%2Fccwc54503.2022.9720869}
\BIBentrySTDinterwordspacing

\bibitem{int2}
M.~J. Hossain~Faruk, H.~Shahriar, and M.~Tasnim, ``Machine learning for
  ransomware: Web-based tutorial for young cybersecurity learner,'' 06 2022.

\bibitem{int3}
L.~Chen, C.-Y. Yang, A.~Paul, and R.~Sahita, ``Towards resilient machine
  learning for ransomware detection,'' 2019.

\bibitem{int4}
A.~Abiola and M.~Marhusin, ``Signature-based malware detection using sequences
  of n-grams,'' \emph{International Journal of Engineering and
  Technology(UAE)}, vol.~7, 10 2018.

\bibitem{mowri2022application}
R.~A. Mowri, M.~Siddula, and K.~Roy, ``Application of explainable machine
  learning in detecting and classifying ransomware families based on api call
  analysis,'' 2022.

\bibitem{lr2}
A.~Vehabovic, N.~Ghani, E.~Bou-Harb, J.~Crichigno, and A.~Yayimli, ``Ransomware
  detection and classification strategies,'' 2023.

\bibitem{lr3}
K.~Lee, J.~Lee, S.~Lee, and K.~Yim, ``Effective ransomware detection using
  entropy estimation of files for cloud services,'' \emph{Sensors}, vol.~23, p.
  3023, 03 2023.

\bibitem{lr1}
T.~Weikert, S.~Basa, and R.~Singh, ``Singh et al-2019-scientific reports,'' 02
  2019.

\bibitem{lr4}
B.~Marais, T.~Quertier, and S.~Morucci, ``Ai-based malware and ransomware
  detection models,'' 2022.

\bibitem{lr5}
S.~Poudyal, K.~P. Subedi, and D.~Dasgupta, ``A framework for analyzing
  ransomware using machine learning,'' in \emph{2018 IEEE Symposium Series on
  Computational Intelligence (SSCI)}, 2018, pp. 1692--1699.

\bibitem{dataset1}
``{Ransomware Detection Dataset},''
  \url{https://www.kaggle.com/datasets/amdj3dax/ransomware-detection-data-set},
  2023, accessed: June, 2023.

\bibitem{dataset2}
``{Ransomware Detection Dataset},''
  \url{https://github.com/securycore/MLRD-Machine-Learning-Ransomware-Detection},
  2023, accessed: June, 2023.

\bibitem{yang2020}
X.~Yang and J.~Ding, ``A computational framework for iceberg and ship
  discrimination: Case study on kaggle competition,'' \emph{IEEE Access},
  vol.~8, pp. 82\,320--82\,327, 2020.

\bibitem{xgb}
``{What is XGBoost?}'' \url{
  https://www.nvidia.com/en-us/glossary/data-science/xgboost/}, 2023, accessed:
  June, 2023.

\bibitem{Liu2012NewML}
Y.~Liu, Y.~Wang, and J.~Zhang, ``New machine learning algorithm: Random
  forest,'' in \emph{International Conference on Information Computing and
  Applications}, 2012.

\bibitem{rf}
``{What is random forest?}'' \url{https://www.ibm.com/topics/random-forest},
  2023, accessed: June, 2023.

\bibitem{other}
F.~D. Gaspari, D.~Hitaj, G.~Pagnotta, L.~D. Carli, and L.~V. Mancini,
  ``Reliable detection of compressed and encrypted data,'' 2021.

\bibitem{sl}
``{scikit-learn: Machine Learning in Python},''
  \url{https://scikit-learn.org/stable/}, 2023, accessed: June, 2023.

\bibitem{rica2022}
R.~M.~A. Molina, S.~Torabi, K.~Sarieddine, E.~Bou-Harb, N.~Bouguila, and
  C.~Assi, ``On ransomware family attribution using pre-attack paranoia
  activities,'' \emph{IEEE Transactions on Network and Service Management},
  vol.~19, no.~1, pp. 19--36, 2022.

\bibitem{Bae2020RansomwareDU}
\BIBentryALTinterwordspacing
S.~I. Bae, G.~B. Lee, and E.~G. Im, ``Ransomware detection using machine
  learning algorithms,'' \emph{Concurrency and Computation: Practice and
  Experience}, vol.~32, 2020. [Online]. Available:
  \url{https://api.semanticscholar.org/CorpusID:198358416}
\BIBentrySTDinterwordspacing

\bibitem{10.1007/s11277-020-07166-9}
\BIBentryALTinterwordspacing
J.~Hwang, J.~Kim, S.~Lee, and K.~Kim, ``Two-stage ransomware detection using
  dynamic analysis and machine learning techniques,'' \emph{Wirel. Pers.
  Commun.}, vol. 112, no.~4, p. 2597–2609, jun 2020. [Online]. Available:
  \url{https://doi.org/10.1007/s11277-020-07166-9}
\BIBentrySTDinterwordspacing

\end{thebibliography}

\end{document}